# UV-SERS monitoring of plasmons photodegradation of biomolecules on Aluminum platforms decorated with Rhodium nanoparticles

Yanqiu Zou[a], Luca Mattarozzi[b], Huaizhou Jin[c], Qifei Ma[d], Sandro Cattarin[b], Shukun Weng[e], Ali Douaki[e], German Lanzavecchia[e], Karol Kołątaj[f], Nicco Corduri [f], Ben Johns[g], Nicolò Maccaferri[g], Guillermo Acuna[f], Zhenrong Zheng[a*], Shangzhong Jin[d*] and Denis Garoli[d,e,h*]

In the search for novel nanostructured materials for UV plasmonics a limited number of choices can be done. Materials such as aluminum, rhodium, gallium and few others can be used. One of the most interesting application for UV plasmonics is Surface Enhanced Raman Spectroscopy. It can be extended to this spectral range to explore spectral properties of biomolecules that have only a small cross section in the visible spectral range. We have recently reported on a functional substrates based on nanoporous aluminum decorated with rhodium nanoparticles. This system showed an interesting behavior for UV excitation at 266 nm, with an unexpected decreasing Raman intensity for increasing rhodium nanoparticles concentrations. We proposed that this effect can be due to the difficult access to the hot spots for the molecules deposited via thermal evaporation. Here we extend this study exploring the performance of the system at another UV excitation wavelengths (325 nm) reporting on experimental results obtained using a deposition process that can bring the molecules at the hot-spots in a more efficient way. Extensive spectroscopic acquisitions, combined with 3D maps, allow to shade a more clear view on the performance of this plasmonic platform. In particular, the photodegration and the potential oxidation of biomolecules driven by the hot-electron/hot-holes produced by the rhodium nanoparticles will be reported.

## Introduction

Surface plasmon resonances find extensive applications in multiple fields such as metal-enhanced fluorescence (MEF), photocatalysis, optical trapping, heating, energy harvesting and Surface Enhanced Raman Spectroscopy (SERS).[1,2] This latter, in particular, has been the subject of intensive research that enabled to demonstrate, during the most recent years, the possibility to obtain Raman spectra from single molecules and to investigate, with a temporal resolution of few milliseconds, the molecular dynamics in chemical reactions.[1,2] Nowadays, SERS platforms are still based on noble metals (Au and Ag) and the application is typically limited to the visible/near-infrared (Vis/NIR) spectral range. As well-known, metals such as aluminum, magnesium, rhodium and gallium can be used to prepare plasmonic platforms working in the spectral range of the UV and deep-UV wavelengths[3,4]. The excitation of biomolecules using UV radiation is particularly interesting for Raman spectroscopy. In fact, most biomolecules have small Raman cross sections in the visible and NIR regions,[5,6] and the use of higher energy to excite them can increase the detection limit due to the presence of electronic resonances[7–11]. Aluminum (Al) is the most extensively explored material for UV plasmonics due to its excellent electrical properties[12], which facilitate significant resonance in the UV range. In order to apply aluminum's plasmonic characteristics for SERS, very small metallic features/nanostructures must be prepared by means of advanced lithographic techniques or by means of chemical synthesis of nanoparticles [13,14,15]. As an alternative approach to produce metals with nanometer features, the preparation of Al films as nanoporous material (NPM) [16–19] has been recently reported with the experimental demonstration of efficient UV-MEF and UV-SERS[6,20,21]. Porous Al structures can be prepared following different strategies[6,22].

Another intriguing approach towards advanced plasmonic platforms consists in the combination of two or more plasmonic materials in the same system[23,24]. In UV-SERS, in particular, the use of rhodium (Rh)[7,25–28] combined with Al could represent an interesting platform to better comprehend how to produce strongly localized electromagnetic fields in the UV spectral range. So, porous Al can be decorated with Rh through a galvanic displacement (GD) reaction, which allows the reproducible preparation of metallic nanoparticles of a more noble metal spontaneously, at open circuit, with the of use of simple chemical apparatus[29].

For this reason, we recently reported on the facile and low-cost fabrication of porous Al substrates decorated with Rh nanoparticles (NPs)[30]. Important to be mentioned, the current main limitation for Rh use in the preparation of plasmonic devices is the very high material cost. The methodology proposed here allows a significant coverage by Rh NPs using minimal Rh amounts. Studying this system for UV-SERS (probing adenine with an excitation laser at 266nm), we observed two interesting phenomena: (i) increasing the density of Rh NPs over the porous Al substrate leads to a decrease in the SERS enhancement; (ii) the Rh NPs slow down the UV-induced photodegradation effect observed during long acquisition time.

In this paper, we extended the investigations on this plasmonic platform performing additional SERS analyses using 325 nm as excitation wavelength and probing two biomolecules, i.e. adenine and bovine serum albumin (BSA), drop-casted from water-based solutions. In particular, adenine is a major contributor to the SERS features of DNA oligomers and strands[31], therefore it is still important to investigate SERS processes on this molecule, in particular in the UV spectral range. We will discuss on the photodegradation process induced by the UV exposure considering also the potential effect of progressive oxidation of adenine to azupurine as previously demonstrated in photoelectrochemical experiments[32]. The progressive oxidation of adenine resulted to be more pronounced for higher Rh NPs contents, suggesting a potential role of the hot-electrons and hot-holes generated by the particles excited at 325 nm.

In addition to these analyses, we will report on the excitation of BSA at long UV wavelengths (out of molecular resonance) discussing both the ability of the platform to detect the main Raman peaks and the photostability of the molecule after multiple

UV exposures. As well-known, serum albumin is the most abundant protein in vertebrate blood and plays an important role as a carrier and in interacting with external molecules and materials. In the study of this protein, deep UV (around 200 nm)[33] was typically used for resonant Raman excitation, enabling the selective enhancement of the signals of amide vibrations occurs, enabling the characterization of the protein secondary structure.

## Results and Discussion

To perform the UV-SERS experiments, we used a new set of samples prepared following a procedure similar to the one reported in our recent publication[30]. Table 1 reports the conditions of GD procedures on Al substrates.

**Table 1** Experimental conditions of sample preparation

| Sample ID | GD bath: | $t_{GD}$ min | Surface Rh coverage % |
|---|---|---|---|
| a | - | 0 | 0 |
| b | 0.5 mM $Na_3RhCl_6$ + 0.09 M NaCl | 2 | 6.46 |
| c | 0.5 mM $Na_3RhCl_6$ + 0.09 M NaCl | 4 | 11.49 |

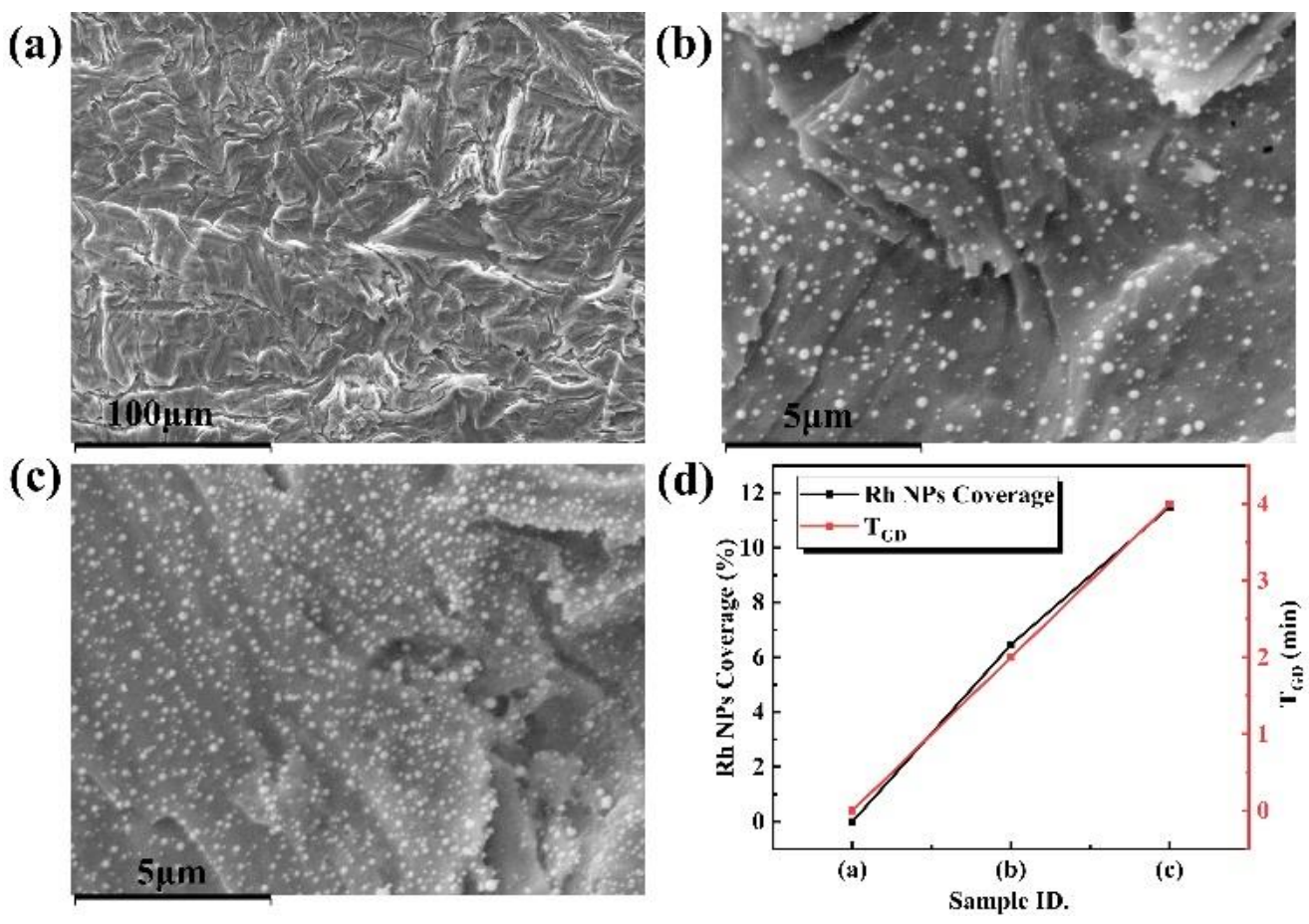

**Fig. 1.** Details on the prepared samples. SEM images of a) untreated rough Al (Sample *a*, no GD) and b-c) Al samples submitted to galvanic displacement in 0.5 mM $Na_3RhCl_6$, pH 2.0, for different times: b) $t_{GD}$ = 2 min (Sample *b*); c) $t_{GD}$ = 4 min (Sample *c*), magnified detail. Line chart of d) Rh NPs coverage per unit area for samples (a-c) at corresponding galvanic displacement times.

Fig. 1(a-c) illustrates SEM micrographs obtained of samples a-c, while Fig. 1(d) reports the coverage of Rh NPs with respect to the corresponding galvanic displacement times for each sample. The samples show a strongly roughened Al surface decorated with different concentrations of randomly distributed Rh NPs. With respect to the previous report[30], no mechanical polishing was executed on the Al substrates to avoid surface damages due to mechanical scratching and inclusions of silicon carbide particles used for the polishing. The pristine surface of the Al exposed to $Rh^{3+}$ ions reacted, facilitating the preparation of Rh NPs with a reduced size distribution (see Supporting Information #1). Cross section imaging allows to observe that the NPs present a morphology approaching a hemi-spherical shape. To verify the electromagnetic field configurations that can be produce by such structures we performed numerical simulations by using of Comsol Multiphysics. The results are illustrated in Supporting Information #2 where we considered two “partial-dome” configurations for the NPs on the Al substrate. We find a strongly enhanced electric field around the Rh NPs, particularly near the gap between Rh NPs and the Al substrate. With respect to the previous experiments[30], here two main points were modified, i.e. the procedure to deposit the target analyte (Adenine and BSA[34–36]) and the excitation wavelength used in the Raman spectroscopy (325 nm). In particular, while in the previous case the samples were coated with the probe molecules by means of physical vapor deposition, here we tested molecules in solution. By using this approach, we expect to achieve a more efficient interaction between the molecules and the gap between the Rh NPs and the Al substrate where the electromagnetic field is more localized[30].

We focused our experiments on the investigation of the photodegradation processes of adenine and BSA due to successive UV laser exposures. In order to do that, we first measured all the samples functionalized with adenine (using a concentration of 0.5 mM). We performed six successive scans over an area of 2x2 $mm^2$. Adenine spectra obtained from all the samples are shown in Fig. 2, showing clearly distinguishable bands that correspond to the molecular vibrations of adenine as reported previously in resonant Raman experiments with adenine and related compounds, as well as in pre-resonant SERS spectra on Rh composite materials[37–41]. Fig. 2(a) illustrates a direct comparison between the samples after a first single laser scan, while Fig. 2(b) depicted the maximum intensity around 731 $cm^{-1}$ and 1331 $cm^{-1}$ for the tested samples. The position of the obtained spectral peaks exhibited shifts. We have provided an explanation in Supporting Information #3 Figure S4(a), Figure S5(a) and (d). Fig. 2(c-d) report the relative 3D time-dependent UV-SERS spectra and mean spectra of adenine across six consecutive scans on sample (a), respectively, clearly showing the photodegradation due to successive UV exposures. We also presented the average UV-SERS spectra of adenine from six consecutive scans on sample (b) and (c) in Supporting Information #3 Figure S4(c) and S4(d), respectively.

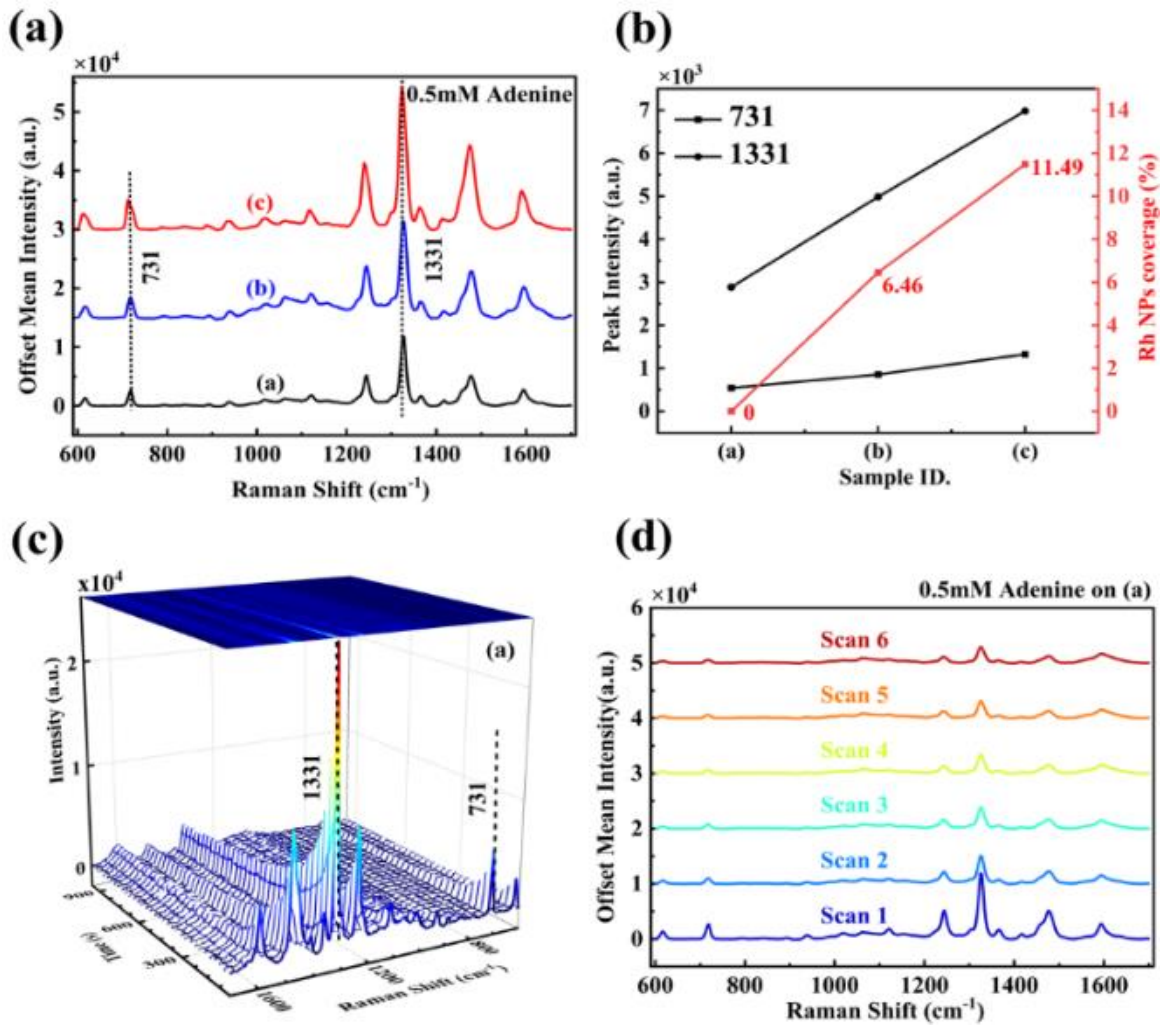

**Fig. 2.** Adenine UV-SERS from Al-Rh NPs samples. (a) Comparison between the first scan obtained from the three tested samples; (b) The maximum values around 731$cm^{-1}$ and 1331$cm^{-1}$ extracted from the spectra of **Error! Reference source not found.**(a); (c) 3D time-depended SERS spectra and mean SERS spectra a cross six consecutive scans of 0.5mM adenine on sample (a), respectively.

As a first important observation, the intensities of the UV-SERS spectrum resulted to increase with respect to the increasing Rh contents on the samples. Fig. 2(b) shows how an increase in Rh NPs% is generally positively correlated with SERS signal intensities at both wavenumbers, with the Pearson correlation coefficient being slightly lower at 1331 $cm^{-1}$ ($r = 0.75$) compared to 731 $cm^{-1}$ ($r = 0.81$). Specifically, sample c (Rh NPs coverage % = 11.49%) exhibited the highest SERS intensities at both wavenumbers, while sample a without GD treatment, served as a control and demonstrated the lowest SERS signal intensities. This is partially in disagreement with our previous results[30] where the Rh NPs seemed to reduce the intensity of the Raman peaks. As already reported, this was due to the not efficient interaction between the plasmonic hot-spots generated by the NPs and the physical evaporated probe molecules. Here, on the contrary, the molecules can interact uniformly with the whole sample thanks to the functionalization method used.

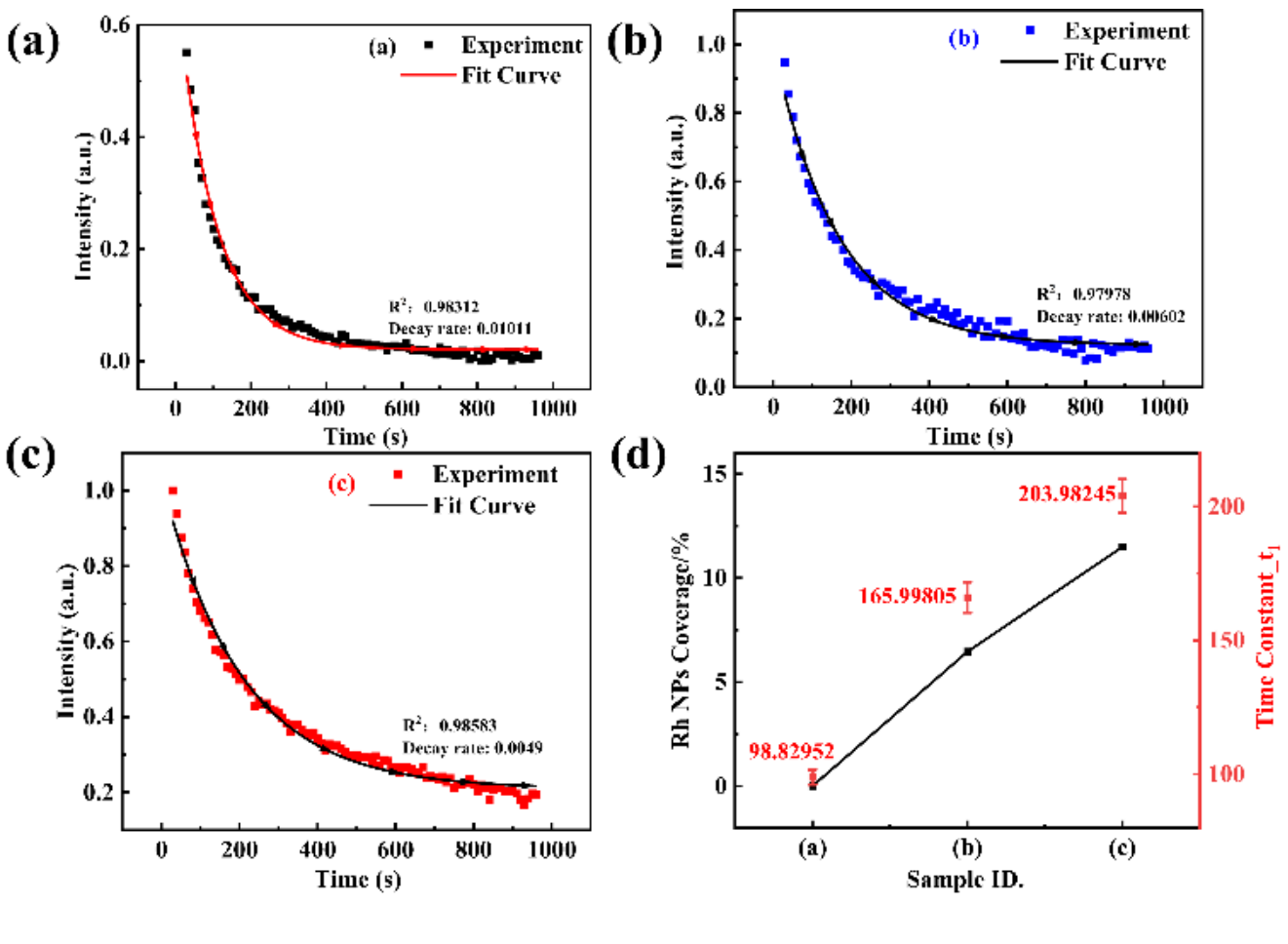

**Fig. 3.** (a-c) The experimental data and corresponding photodegradation fitting curves of peak intensity at around 1331 $cm^{-1}$ for adenine as a function of collection time for different samples a-c respectively. (d) Time constants of sample (a-c) with varying Rhodium Nanoparticles coverage %.

In all the measurements, the two most characteristic SERS peaks of adenine at around 731 and 1331 $cm^{-1}$ were identified[32,42]. However, there are several other strong bands that appear between 1000 and 1500 $cm^{-1}$, such as 1048, 1423, and 1482 $cm^{-1}$. The first observation of this kind of SERS spectrum was reported for adenosine on a roughened silver electrode at a positive potential in 1985.[43] We have also conducted the SERS spectra of several derivatives of adenine in Supporting Information #4. As shown in Figure S6, the results indicated that we can observe vibrational band for adenine and its derivatives $NAD^+$, ATP and adenosine within the wavenumber of 1000 and 1100 $cm^{-1}$ as similar to the literature[32], but the peak also shows different shifting to around 1070$cm^{-1}$.

As a first step in the photodegradation analyses, we monitored the intensity of the peak at 1331 $cm^{-1}$ during six mapping scans over the different samples. The mode at 1331 $cm^{-1}$ corresponds to a skeletal vibration of the pyrimidine ring[44] and its intensity could therefore be indicative of decomposition of the molecule. The decrease in intensity of this band could indicate dissociation of skeletal purine ring bonds, and/or a re-orientation of the molecule on the nanostructure.

Combining the Rh NPs coverage in Table 1 and the time constant under 1331 $cm^{-1}$ respected to the Rh NPs content at Fig. 3(d), we can conclude that Rh incorporation generally improves photostability, as evidenced by increased $t_1$ values in samples with higher Rh NPs content. Specifically, the standard deviations at 1331 $cm^{-1}$ and 1482 $cm^{-1}$ are both less than 10 indicates the fitting results are in good agreement with Eq. (1). Sample (c) (11.49% Rh NPs coverage) showed the highest $t_1$ at 1331 and 1482 $cm^{-1}$, while Sample a, with no Rh content, exhibited the lowest $t_1$values, suggesting that Rh contributes significantly to photostability. However, for the photodegradation constants of adenine at the fitted peaks of 1048, 1423 and 1067 $cm^{-1}$, samples containing Rh NPs generally exhibit large standard error values and even anomalies. This indicates that using the photodegradation Eq. (1) alone is no longer sufficient to meet the fitting convergence conditions. Moreover, these peaks are roughly classified as "unusual peaks" as defined in Reference[30]. For unusual peaks, further analyses are necessary.

**Table 1.** Calculate the photodegradation Time Constant ($t_1$) for the three samples across various wavenumbers

| ID. | Time constant ($t_1$) under different wavelength | | | |
|---|---|---|---|---|
| | 1048 $cm^{-1}$ | 1067 $cm^{-1}$ | 1423 $cm^{-1}$ | 1482 $cm^{-1}$ |
| (a) | 75.3 ± 14.7 (s) | 69.3 ± 8.3 (s) | 105.1 ± 8.8 (s) | 101.6 ± 3.6 (s) |
| (b) | 90.5 ± 25.9 (s) | 96.4 ± 21.5 (s) | 118.3 ± 18.7 (s) | 137.3 ± 8.4 (s) |
| (c) | 543.2±148.9 (s) | -485.2 ± 279 (s) | 226.7 ± 23.2 (s) | 221.3 ± 7.7 (s) |

In order to better understand the potential oxidation reaction from adenine to azupurine, additional data analyses on the collected spectra were performed calculating the area ratio between different spectral ranges. The calculated results by Python were shown in Supporting Information # 6 Figure S8, which demonstrate that peak areas in the [700, 745] $cm^{-1}$ range initially decrease and then stabilize, the [1285, 1352] $cm^{-1}$ range exhibits a gradual reduction slowed by Rh nanoparticles, and the [960, 1200] $cm^{-1}$ range generally decreases with increasing scans, accompanied by ratio analyses indicating consistent normal peak ratios and varying relative changes in other peak ratios.

To investigate whether Rh NPs have an oxidation effect on adenine and alter the "unusual peak" of adenine around ranges of 1000 to 1100 $cm^{-1}$ (According to the literature, silver colloid-induced visible adenine SERS exhibits oxidation effect on characteristic peaks around 1048 $cm^{-1}$)[32] , we performed multiple peak deconvolution on the [960,1200] $cm^{-1}$ range and single peak fitting on the [700,745] $cm^{-1}$ range by applying Gaussian functions, using the formula[47]:

$$y = y_0 + \frac{Ae^{\frac{-4\ln(2)(x-x_c)^2}{w^2}}}{w\sqrt{\frac{\pi}{4\ln(2)}}} \qquad (1)$$

where $x_c$ is the center peak wavenumber, setting the initial value $y_0$ = 0. Figure S9 in Supporting Information #7 displays the fit curve of sample (a) as an example to illustrate our reporting method. Using Eq. (2), we performed the peak fitting and calculated the corresponding area and relative full width at half maximum (FWHM). Since the area corresponding to the wavenumber range of [700-745] $cm^{-1}$ does not change significantly with an increasing number of scans, we selected the fitting peak at approximately 731 $cm^{-1}$. We then calculated the ratio of the peak area for each fitted peak to the peak area around 731 $cm^{-1}$ obtained from the same spectrum. The calculation results are shown in Supporting Information #7 Figure S10, which report the variation of the peak area corresponding to the fitting peak with the number of scans. The results show the ratio of the fitted peak areas increases as the number of scans increases. When combined with the Rh NPs coverage percentage in Table 1, we can observe how higher concentrations of Rh NPs seem to promote greater oxidation effects, as evidenced by the steeper slopes observed at around 1075 $cm^{-1}$ fitted peak, which is indicative of enhanced oxidative interactions within the sample matrix, as reflected in the altered SERS peak area ratios within the 1000-1100 $cm^{-1}$ range. It is interesting to note that this oxidation process seems to be not accompanied by the oxidation of carbon species on the samples' surface. In fact, even after multiple laser scans, we did not observed increasing peak intensity around 1600 $cm^{-1}$, where the main carbon oxide vibration is expected[48,49].

UV-SERS experiments using adenine as probe molecule have been reported in several papers[5,6,15,22,30,46]. On the contrary, measurements on proteins were not so extensively discussed[33,50]. Here we test our substrate also as a platform for UV-SERS using BSA, a protein whose SERS spectroscopy has been previously reported[51–53].

We first tested the bulk BSA on silica chip, the corresponding result was reported in Supporting Information #9 Figure S11. Then we measured the Al-Rh NPs substrates using 15.1 µM BSA concentration in water solution and we performed multiple scans over the same area to evaluate the effect of successive UV exposures.

Fig. 4 reports the spectrum obtained after a single scan from the different samples and illustrates the relative changes in the intensity ratios of the peak between 1050 to 1070 cm$^{-1}$ and 1500 to 1700 cm$^{-1}$ in the UV-SERS spectra of BSA across four consecutive scans on different substrates (a, b, c). As can be seen, with respect to adenine, in this case only few peaks can be detected. In particular, the relatively broad peaks at around 1606, 1360 and at 1064 cm$^{-1}$ can be associated to vibration of BSA in agreement with previous reports[33,34,51,50].

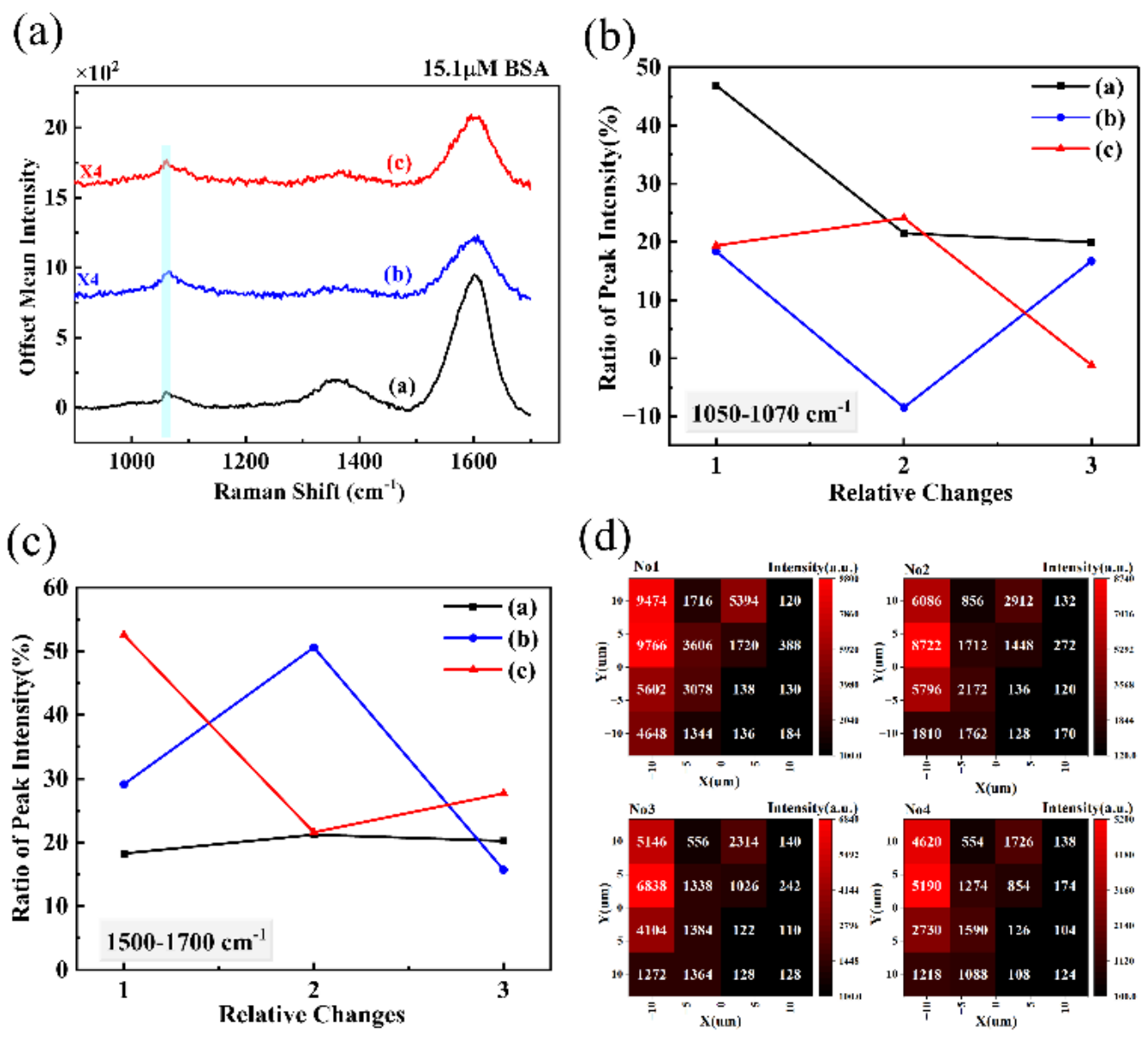

**Fig. 4.** 15.1 µM BSA UV-SERS on different samples a-c (black, blue and red spectra respectively) and peak intensity analysis. (a) Mean Intensity; (b-c) Ratios of peak intensity changes across four scans for Rh-modified Al ranging from 1050 to 1070 cm$^{-1}$ and 1500 to 1700 cm$^{-1}$ respectively; (d) Mapping images of sample a across four scans between the range of 1500 and 1700 cm$^{-1}$.

Important to note, here the excitation laser at 325 nm is not in resonance with the molecule, therefore we were able to detect only a small number of vibrations with respect to other reports where the resonant Raman was obtained using higher frequencies for the laser excitation[50]. In this case it is not easy to assign the peaks position to clear molecular's vibrations. In particular, the peak at 1600 cm$^{-1}$ could be associated to the presence of carbon oxide species on the substrate surface, even if we did not observe an increasing intensity for successive scans and we did not observe this carbon vibrations in the multiple Raman experiments performed with adenine.

Sample (a) (black line in Fig. 4(a)), which does not contain Rh NPs, exhibits the strongest BSA Raman enhanced signal and a broad peak around 1360 cm$^{-1}$. The relative peak intensity between the range of 1050 to 1070 cm$^{-1}$ and 1500 to 1700 cm$^{-1}$ of BSA UV-SERS spectra evaluated across three scan intervals (labeled 'relative changes' as x-axis) for three samples were reported in Fig. 4(b) and (c), respectively. Taking the Rh NPs coverage percentage in Table1 into account, we can conclude that Rh NPs exert some concentration-dependent influences on the vibrational dynamics of the system. Specifically, the 1600 cm$^{-1}$ peak in Fig. 4(c) exhibits a significant initial enhancement in samples with Rh NPs content, especially in sample c, where the intensity ratio reaches 52.55% after the first scan. Subsequently, the intensity ratio decreases and slightly rebounds in the following scans, reflecting rapid structural changes and subsequent reaction dynamics catalysed by Rh NPs. In contrast, the 1060 cm$^{-1}$ peak in Fig. 4(b) continuously decreases in the Rh NPs-free sample a, whereas in samples with Rh NPs, it shows a complex trend. Initially, the intensity ratio may increase due to the formation of specific structures or oxidation products, followed by a decrease caused by excessive oxidation or structural disruption. These results indicate that the coverage of Rh NPs significantly influences the structural changes of BSA, thereby affecting the intensity ratios of peaks in the UV-SERS spectra. This emphasizes the crucial role of Rh NPs in regulating protein structure and SERS signals. To ensure that each scan covered the identical region, the scanning paths alternated between x [-20,20] y [-20,20] and x [20, -20] y [20, -20]. As shown in the Fig. 4(d), the heatmap within range of 1500 to 1700cm$^{-1}$ remain consistent across all four scans, confirming that the same area was repeatedly scanned.

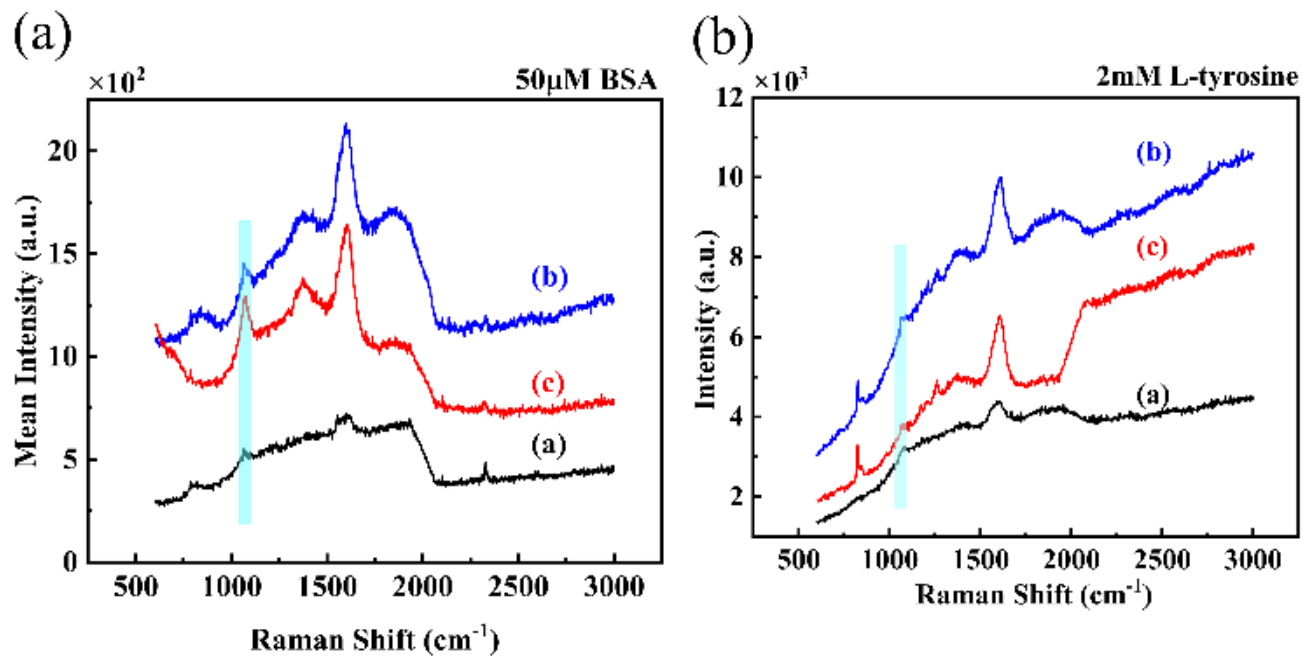

**Fig. 5.** UV-SERS spectra of (a) 50 µM BSA and (b) 2 mM L-tyrosine UV-SERS on samples (a, b, c) after storage for 2.5 months.

We also performed experiments to test 50 µM BSA on the same samples (a, b, c) after storing them for 2.5 months to check the time stability of our substrates, the result was reported in Fig. 5(a). To complement our UV-SERS measurements of BSA, we also examined the SERS spectra of L-tyrosine using the same samples (a, b, c), following the approach of the literature[33], which highlighted the significant role of tyrosine residues in the UVRR spectra of BSA. While we only tested the UV-SERS spectra of 2 mM L-tyrosine dissolved in Milli-Q ultrapure water, result represented in Fig. 5(b). By comparing the UV-SERS spectra of BSA and L-tyrosine in Fig. 5(a) and (b), we found that the spectra of these two substances are very similar. This also indicates that the significant role of tyrosine residues in the UVRR spectra of BSA, as highlighted in the literature, is also applicable to not-resonant UV-SERS.

## Conclusions

In this paper we investigated, in details, the performances, in terms of photostability effects, of an UV-plasmonic platform based on rough Al decorated with different concentrations of Rh NPs. While the combination of a porous Al film with Rh NPs was already proved for applications in UV-SERS[30], the analyses reported here enable a better understanding of the behaviour of this particular plasmonic platform. In particular, using a suitable functionalization strategy to bring the probe molecules all over the sample we verified that Rh NPs can contribute to increase the enhancement in the SERS signal. Not less interesting, we explored the effect of Rh NPs in the photostability under successive UV exposure discussing both oxidation effects and degradation effects in adenine and BSA. The results reported here contribute to better comprehend the UV-SERS phenomenon and can be extended to more efficient configurations such for example nanoparticles on a mirror[54–56], where the Rh NPs can be first functionalized with the probe molecules (for example DNA) and then deposited on an Al film.

## Experimental

### Sample preparation

Alfa Aesar 99.99% Al foil sheets were cut into approximately 1 x 1 cm2, with a thickness of 1.5 mm. The side not subject to galvanic displacement was marked with a mechanical engraver.

After cutting, each sample was first washed with acetone, then with deionized water in an ultrasound bath (5 min) to remove any contamination and dried. Subsequently, the sample was subjected to chemical etching in 5M $H_2SO_4$ at 80 °C (5 min). After the chemical treatment, the square foils were treated with ultrasonics in water for an additional 5 minutes, dried only on the backside (where the marking is present), and deactivated with tape for anodization (3M 8992). The GD processes were carried out using solutions prepared with deionized water (Elga-Veolia Purelab Pulse System, 18 MΩ·cm) and high-purity reagents: NaCl (99.5%, Merck), $Na_3RhCl_6{\cdot}12H_2O$ (Alfa Aesar), and HCl (37%, Merck). To enhance reproducibility - considering the slow equilibration of Rh chlorocomplex speciation[57,58] - the solutions were aged for 48 hours at 60 °C.

The solution, acidified to pH 2 with HCl, was kept at 25 °C and purged with nitrogen for 20 minutes prior to sample immersion.

After the galvanic exchange, the foils were washed in water, dried under a mild $N_2$ flow, and stored under vacuum pending morphological characterization (SEM).

### Sample morphological characterization

Scanning electron microscopy (SEM) images were acquired using a Zeiss SIGMA microscope equipped with a field emission gun, operating under high vacuum and an accelerating voltage of 20 kV.

Image analysis performed using the open-source software ImageJ allowed for the estimation of surface area fraction of Rh NPs covering the Al substrates.

**Materials**

Adenine (99%) and L-Tyrosine (99%) were purchased from J&K-Sci. BSA was purchased from Solarbio Life Science. $NAD^+$(98%) was purchased from Shanghai Shaoyuan Co.Led. ATP (97%) and Adenosine were purchased form Adamas-beta. All solutions were prepared by using Mili-Q ultrapure water as solvent.

**Photostability tests with adenine**

The substrates were dropped with 2μL 0.5mM adenine using micropipette, and tested after naturally drying at room temperature. For each mapping measurements, a small area around 2 × 2 $\mu m^2$ (total 16 points for each scan) was performed. The tests were conducted in six consecutive scans for 0.5mM adenine across the same small area with 10s exposure time and one accumulation for each point. The same laser and acquisition parameters were maintaining the same for three samples. The raw spectra were baseline-corrected and smoothed using LabSpec6 (HORIBA measurement software) before processing averaged spectra and peak deconvolution analysis by Origin.

**Photostability tests with BSA**

Samples (a, b, c) were dropped dropped with 2μL 15.1μM adenine using micropipette, and tested after naturally drying at room temperature. For each mapping measurements, a total area 20 × 20 $\mu m^2$ (total 16 points) with a 6.667μm step size was performed. The tests were conducted in four consecutive scans across the same area with 10s exposure time and one accumulation for each point. The same laser and acquisition parameters were maintaining the same for three samples. The raw spectra were baseline-corrected and smoothed using LabSpec6 (HORIBA measurement software) before processing averaged spectra and peak intensity analysis.

**Raman spectroscopy:**

SERS measurements were performed using a HORIBA LabRAM HR Evolution Raman spectrometer (Horiba Jobin Yvon, Kyoto, Japan) with 40X UV objective and 50X long-focal-length objective (NA = 0.75). The UV-SERS spectra of adenine, BSA and L-Tyrosine were conducted using 325nm laser at 10% power attenuation, 500-hole value and 1800gr/mm grating with 10s exposure time, and one accumulation. Visible Raman spectra of bulk BSA were measured using 633 nm laser at 10% power, 532 nm laser at 1% power and 785nm laser at 10% power, and spectra were acquired with a 600 gr/mm grating, 10s exposure time, and one accumulation. For mapping measurements, the tests were conducted in six consecutive scans (small area total 16 points for each scan) for 0.5mM adenine and four consecutive scans (20 X 20$\mu m^2$, total 16 points, 6.667μm step size) for 15.1μM BSA across the same area. The same laser and acquisition parameters were maintained for each measurement scan. For measuring 50μM BSA, 2mM L-tyrosine and several derivations of adenine on samples stored for 2.5 months, it is necessary to increase the acquisition time to 180 seconds with a single accumulation and 10% laser power. The spectra shown in Fig. 5 were obtained by averaging three randomly collected spectra without smoothing filter and baseline, and the abrupt changes at around 2000 $cm^{-1}$ were due to rating alteration.

**COMSOL Multiphysics modelling**

The electromagnetic field configurations produced by the Rh NPs on Al substrate was simulated using the wave optics module in COMSOL. A plane wave excitation at 325 nm wavelength is used to obtain the near-field distribution around Rh NPs of different sizes in partial-dome configurations (see Supporting Information for details).

## Author Contributions

YZ, HJ, SW and QM performed the Raman measurements and data analysis, LM and SC conceived and performed the films synthesis, AD and GL performed samples characterizations, BJ, and NM the numerical simulations, AD and GL supported in samples preparation, GA, KK, and CN supported in data analysis and paper writing, ZZ, SJ, and DG conceived the experiment and supervised the work. The authors thank the IIT clean room facility.

## Conflicts of interest

There are no conflicts to declare.

## Acknowledgements

The authors thank National Natural Science Foundation of China (No.22202167), National Key, Research and Development Project of China (No. 2023YFF0613603), the European Union under the Horizon 2020 Program, FET-Open: DNA-FAIRYLIGHTS, Grant Agreement 964995, the HORIZON-MSCA-DN-2022: DYNAMO, grant Agreement 101072818. NM and BJ acknowledge the 'Excellence by Choice' Programme at Umeå University funded by Kempestiftelserna (grant no. JCK-2130.3). The authors thank Clean Room Facility of IIT.

## Notes and references

**Supporting Note #1.** *Additional SEM images of Rh NPs prepared via GD*

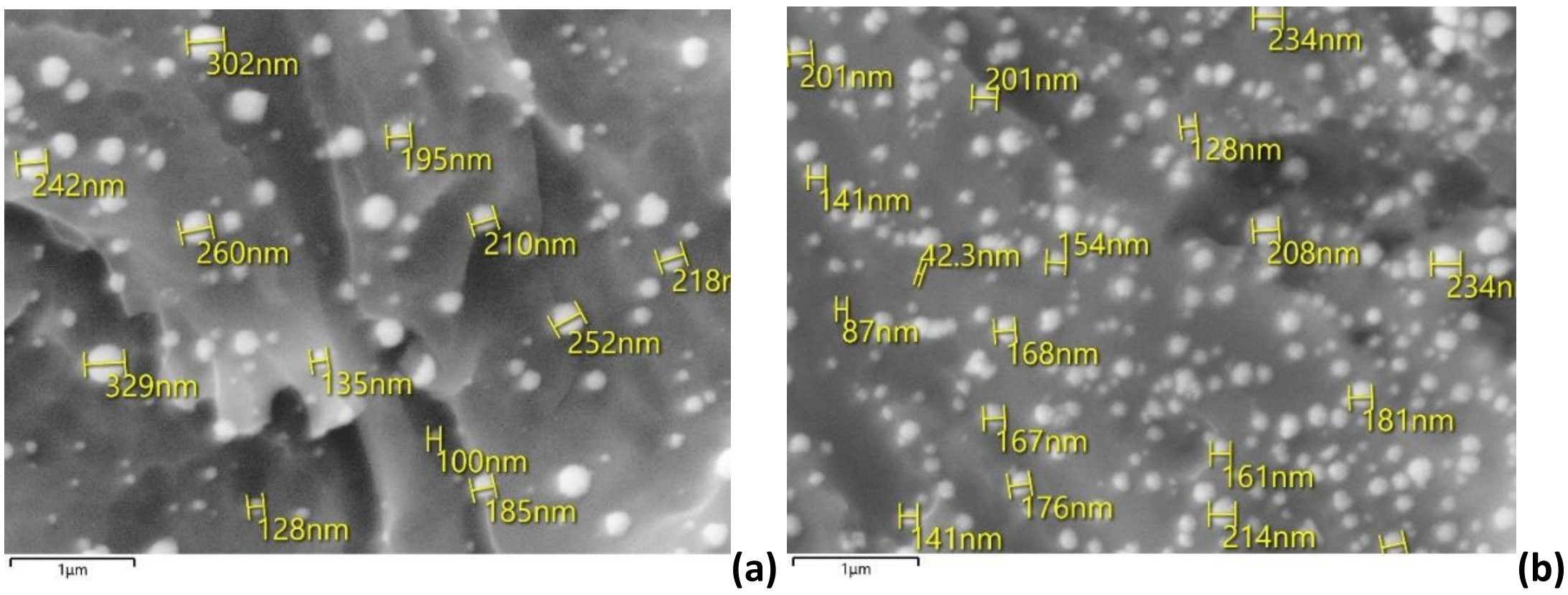

**Fig. S1.** Additional details on the prepared Al-Rh NPs samples. SEM images of a) Al samples submitted to galvanic displacement in 0.5 mM $Na_3RhCl_6$, pH 2.0, for $t_{GD}$ = 2 min (Sample b); b) $t_{GD}$ = 4 min (Sample c), magnified detail with particles' size.

**Supporting Note #2.** *COMSOL simulations of Rhodium nanoparticles on Aluminum substrate*

The electric field distribution around rhodium nanoparticles (Rh NPs) of different sizes placed on aluminum (Al) substrates was simulated using the wave optics module of COMSOL Multiphysics at a free space wavelength of 325 nm. Two geometries were simulated – one where the NP is directly placed on Al, and the other where the NP is placed on the Al substrate with a thin coating of aluminum oxide. Further, the NP was considered to be either a half-dome or a partial (two-thirds) dome. The simulated electric field strengths of the half-dome Rh NP (normalized to the field strength of the incident plane wave) are shown in Fig. S2. The results for the NP on Al substrate are shown in the top panel (Fig. S2a-c) for NP radii of 50 nm, 75 nm, and 100 nm. The corresponding results for the NP on aluminum oxide coated Al are shown in Fig. S2d-f.

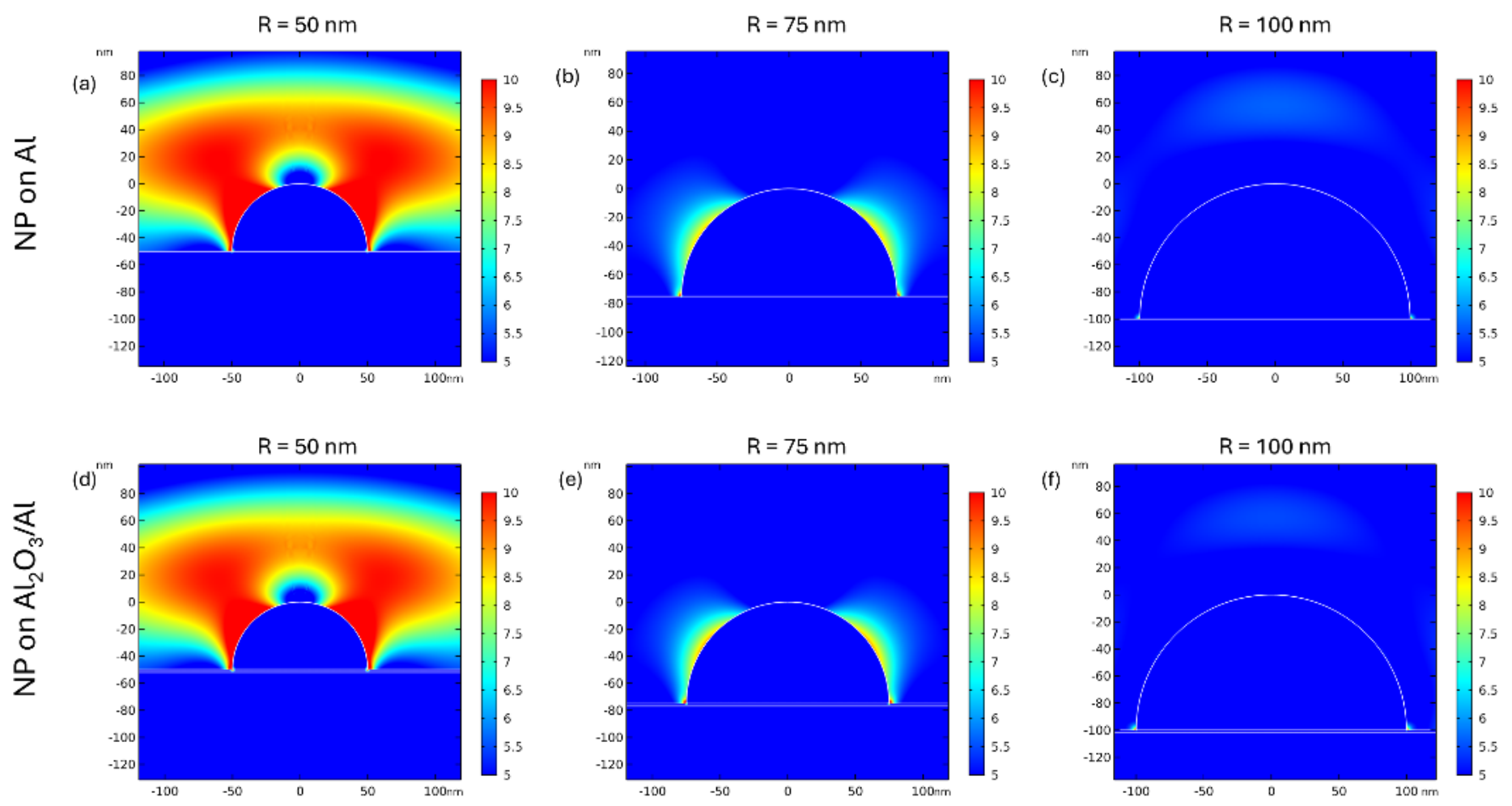

**Figure S2.** Rh half-dome NPs on Al (top row) and $Al_2O_3$/Al (bottom row). The radii of the NPs are indicated above each panel. The thickness of the $Al_2O_3$ layer is 2 nm. The color bar range for $E/E_0$ is fixed at 5-10 for ease of comparison across panels.

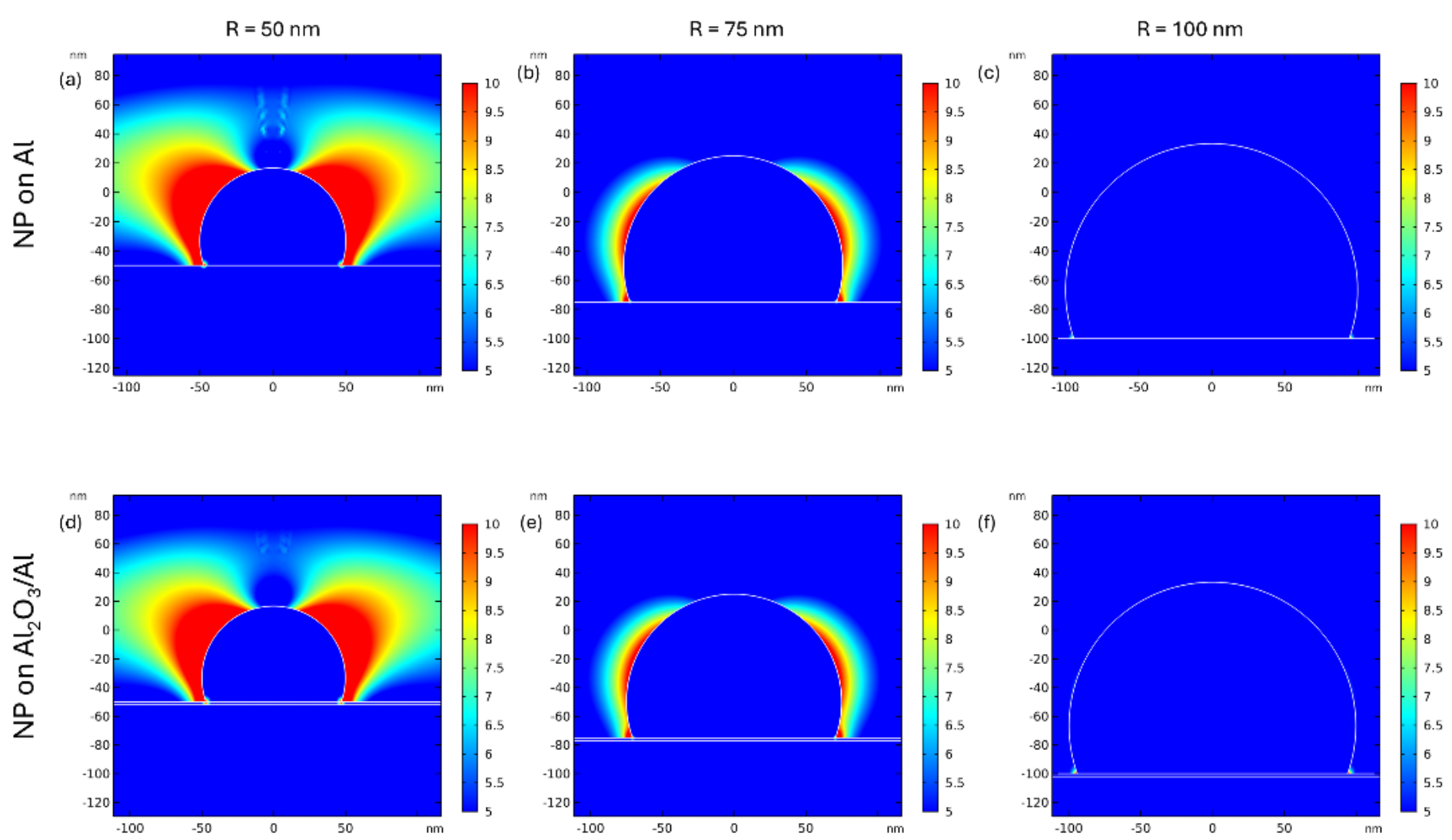

**Figure S3.** Rh partial-dome NPs on Al (top row) and $Al_2O_3$/Al (bottom row). The NP height is two-thirds of its diameter. All other parameters are the same as in Figure S1.

Fig. S3 shows the results for the partial-dome Rh NPs with and without an $Al_2O_3$ coating on the Al substrate. In this case, a sphere with height equal to two-thirds of the diameter is considered. In both the half-dome and the partial-dome on substrate models, the strongest fields are in the region where the dome meets the substrate (for example, see R = 100 nm in Figure S2 and Figure S3). To check for numerical artifacts at this corner, a maximum mesh size at the edge point was set to be 0.1 nm. However, no visible difference is found between this and the results calculated with a maximum mesh size of 1 nm.

**Supporting Note #3.** *Raman Spectrometer calibration correction*

It should be noted that our instrument lacks automatic calibration for ultraviolet (UV) system, resulting in shifts in the spectra obtained during preliminary tests (the data in the main test), we plotted the spectrum from a single point selected from the first mapping scan of 0.5mM adenine on three samples, as shown in Fig. S4(a). The main peaks of adenine are observed at 716.273 and 1325.45 $cm^{-1}$ showed several wavelength shifts (7-8 $cm^{-1}$). Fig. S4(b) presents an enlarged view of the 800–1200 $cm^{-1}$ wavenumber region from Fig. S4a after baseline correction. A prominent Raman peak is observed around 1064 $cm^{-1}$. Fig. S4c and S4d supplement the main text by displaying the average SERS spectra of 0.5 mM adenine obtained from six scans on samples (b) and (c), respectively. The figures illustrate that the average spectral intensity of adenine gradually decreases and stabilizes with an increasing number of scans.

After 2.5 month later, we calibrated the system with a silicon wafer to 520.586 $cm^{-1}$, as shown in Fig. S5a, and the raw spectrum of adenine obtained from subsequent tests across three samples are presented in Fig. S5b-d.

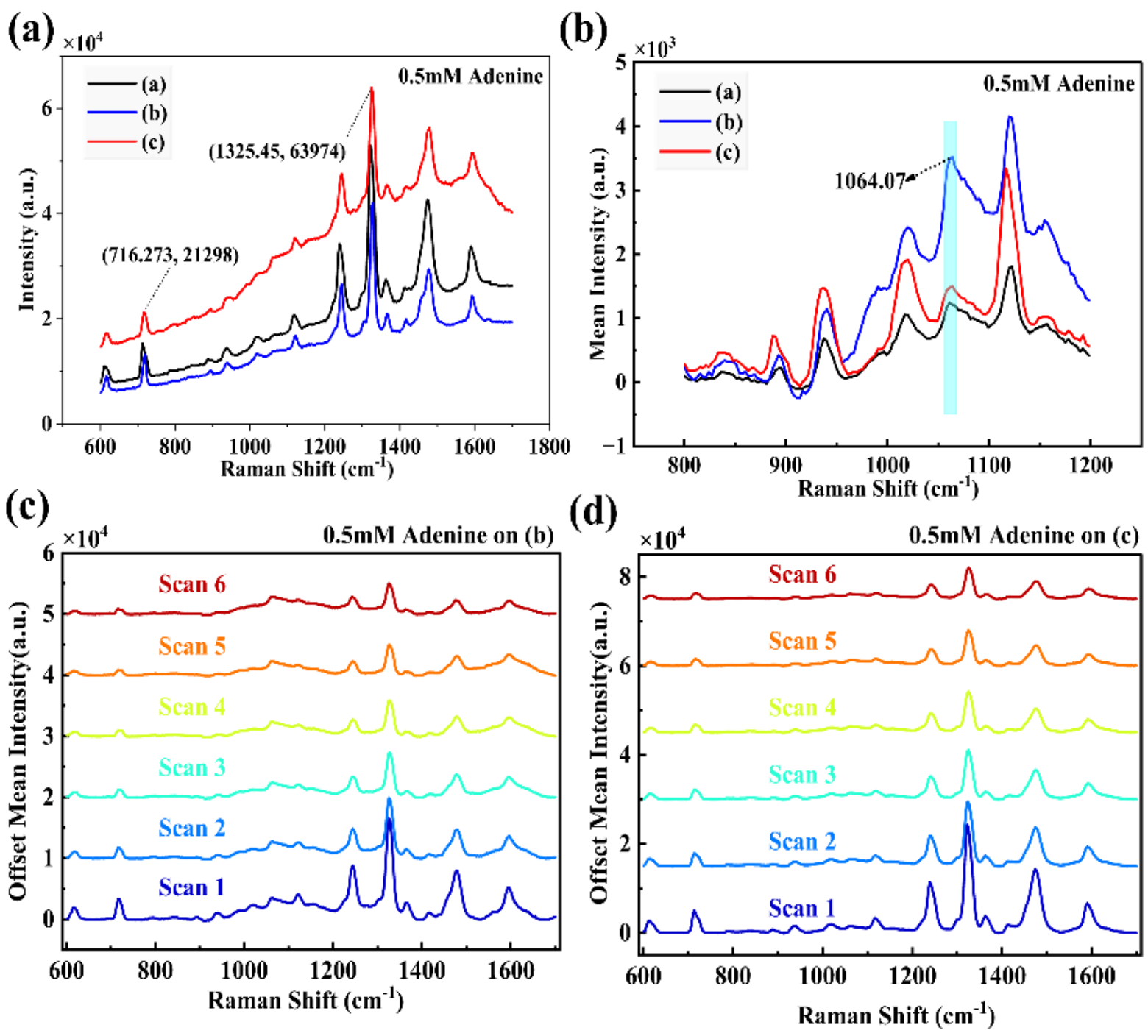

**Figure S4.** Adenine spectra obtained from a single point selected during the first mapping scan across three samples (a: black line; b: blue line; c: red line)

Compared to Fig. S5, the SERS intensity of adenine has significantly decreased, while the main Raman peaks of adenine remain clearly discernible. However, the main peaks of adenine are primarily observed at around 722.996 $cm^{-1}$ and 1329.01 $cm^{-1}$ for our instrument, while also showed several wavelength shifts compared to the peak positions corresponding to adenine spectra reported in the literature (https://doi.org/10.1021/acs.jpclett.3c00619). In the main text, we all labeled the adenine peaks at 731 $cm^{-1}$ and 1331 $cm^{-1}$.

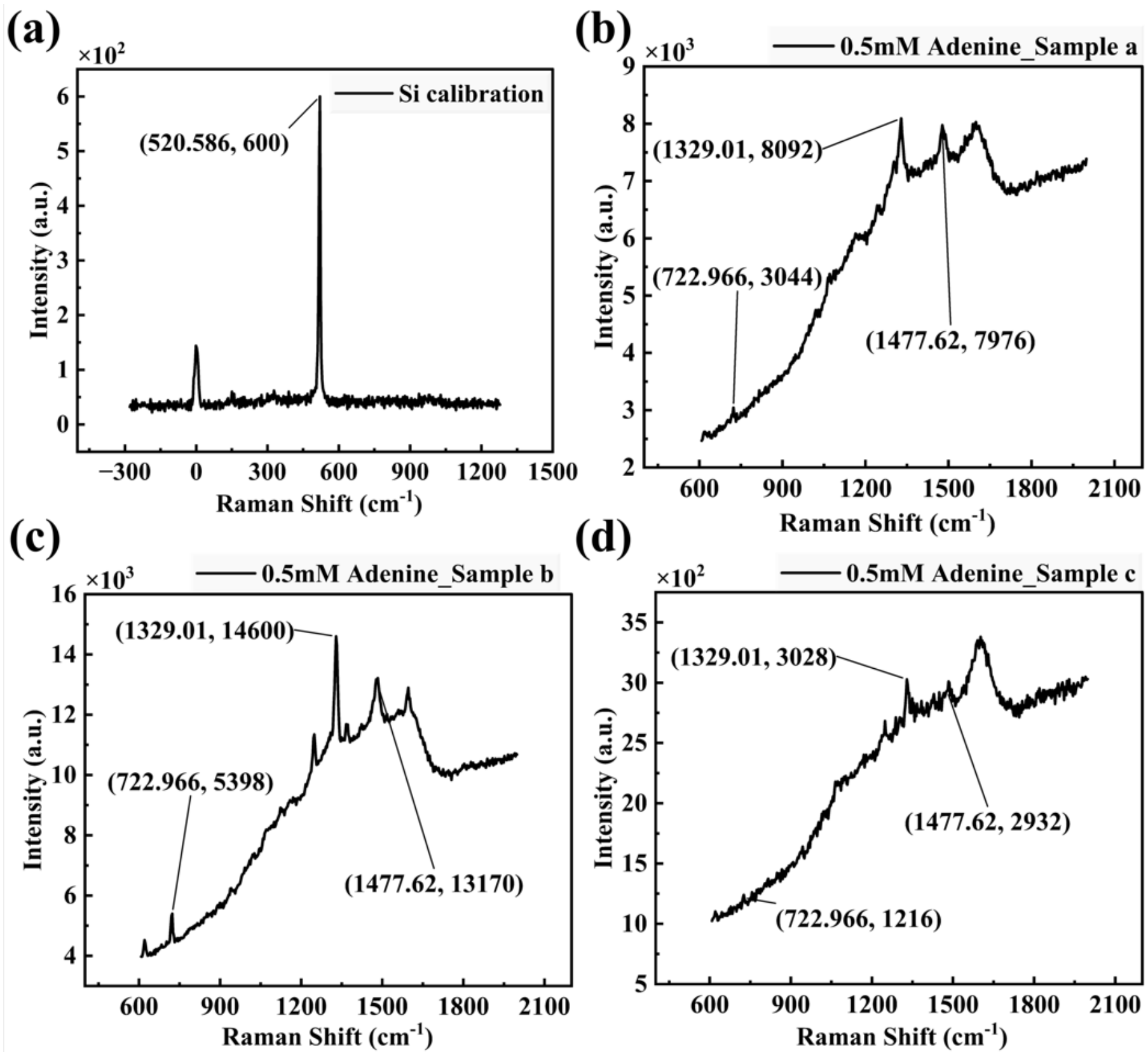

**Figure S5.** Original spectra obtain by our system. (a) Silica Raman spectrum for system calibration; (b-d) the Adenine SERS spectra across different samples (a-c) respectively, labeled with the specific Raman peak position after Silica calibration.

**Supporting Note #4.** *UV-SERS Spectra of several derivatives of Adenine*

As shown in Fig. S6, we tested the SERS spectra of several derivatives of adenine, including 10 mM adenosine, $NAD^+$, and ATP, and compared them with the SERS spectrum of 0.5 mM adenine. Due to the samples being stored for over two and a half periods before testing, the SERS signal intensities of the various substances obtained were relatively low. However, unlike in the literature (https://doi.org/10.1021/acs.jpclett.3c00619), our systematically tested spectra exhibited a prominent peak around 1070 cm$^{-1}$ within the wavenumber range of 1000–1100 cm$^{-1}$.

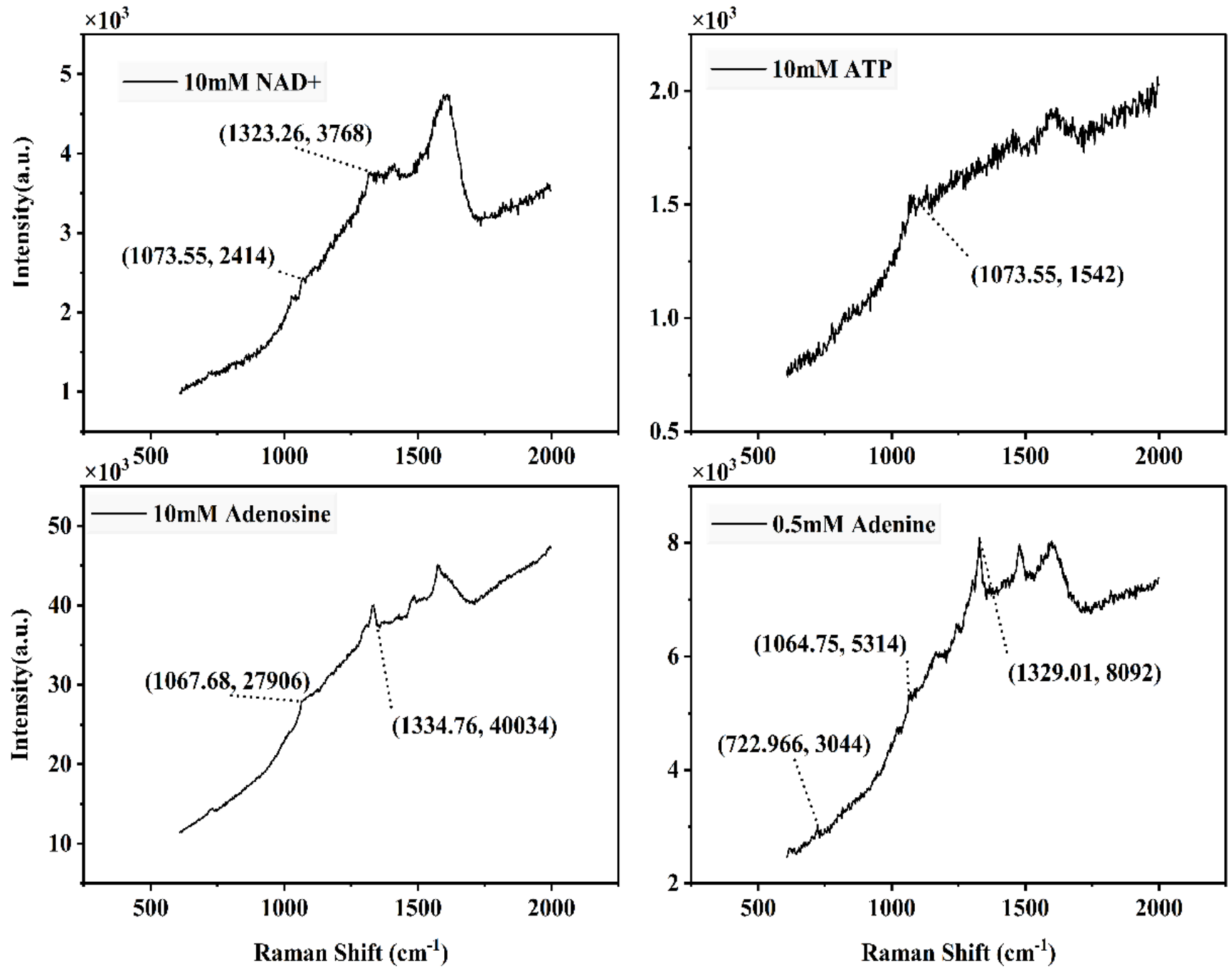

**Figure S6.** Raw SERS spectra of (a) 10mM NAD+, (b) 10mM ATP, (c) 10mM adenosine and (d) 0.5mM adenine on sample a, respectively. The corresponding peak positions and intensities are marked in the spectra.

**Supporting Note #5**. *Photodegradation analyses of adenine*

We apply Eq. (1) in the main text to fit the peak intensities at around 1048 $cm^{-1}$ ,1067 $cm^{-1}$, 1417 $cm^{-1}$ and 1477 $cm^{-1}$ for adenine as a function of collection time for the three samples, the photodegradation fitting curves of the relative intensity are show in Fig. S7. For example, the figure shows the variation of the relative peak intensity at 1048 $cm^{-1}$ on the different substrates over multiple scans (time). It can be observed that the intensity changes at 1048 $cm^{-1}$ differ from those at 1331 $cm^{-1}$. The trend differs from that of the regular peaks. As seen in the figures, the root mean square errors (RMSE) of the fittings are all below 0.6 or even smaller, making accurate fitting challenging. This necessitates further analysis of the corresponding spectral bands and peak areas.

Fig. S7d presents the experimental data of peak intensity variations over time at the adenine characteristic peak of 1482 $cm^{-1}$ for different samples, along with the fitted curves derived from Eq. (1). The results indicate that the root mean square error (RMSE) of the fitting at this wavenumber is consistently greater than 0.9, demonstrating that the fit aligns well with fitting Eq. (1). In contrast, for the wavenumbers corresponding to Fig. S7a-c, the RMSE values of the fittings are generally smaller, and the fitting coefficients at 1048 $cm^{-1}$ and 1067 $cm^{-1}$ are even below 0.4. This suggests that the fittings did not converge, indicating that a single photodegradation Eq. (1) is inadequate for analysis at these wavenumbers.

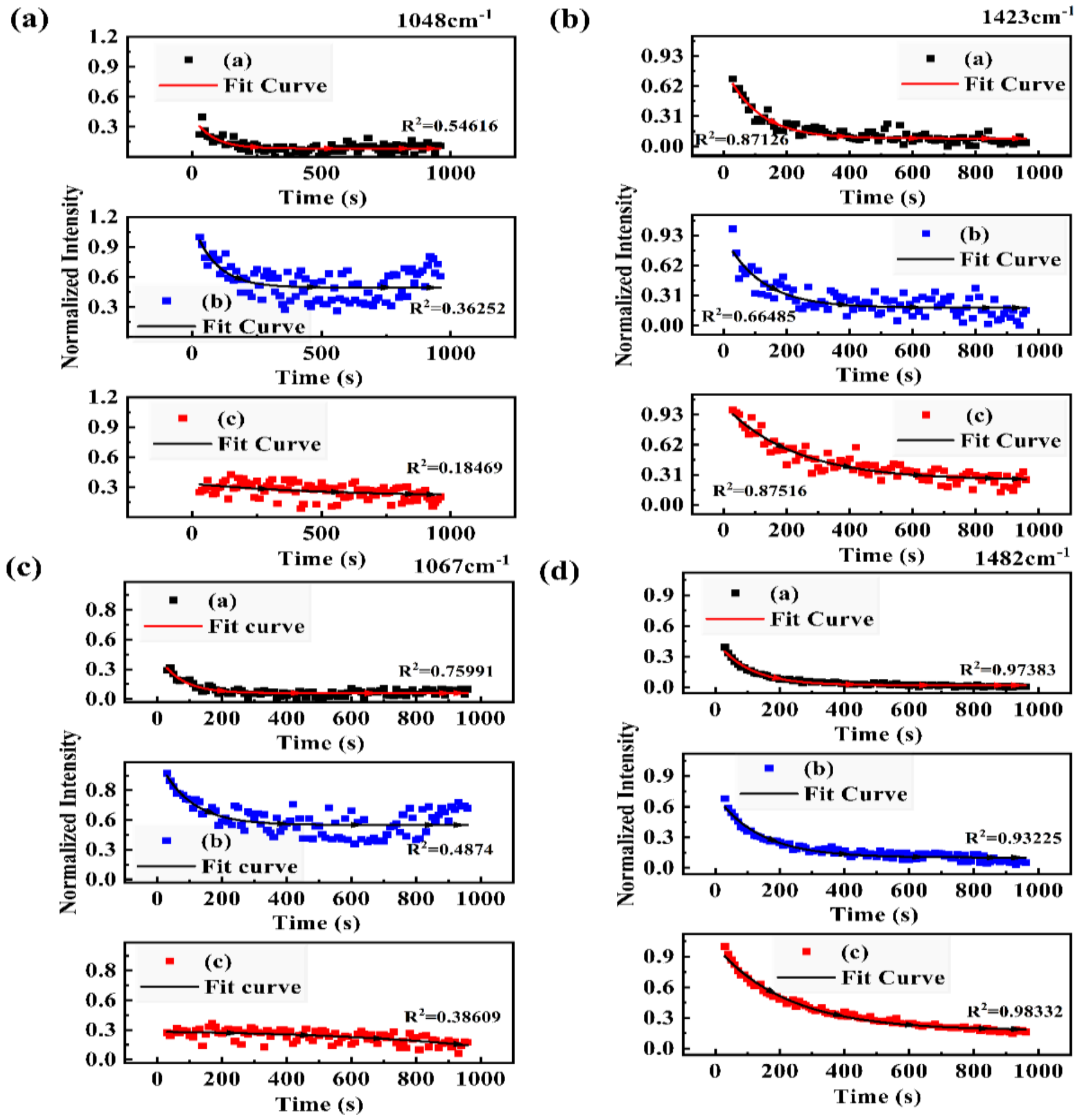

**Figure S7.** Experimental data and corresponding fit curves of 0.5 mM adenine with varying time across three samples under different wavenumbers of (a) 1048 $cm^{-1}$, (b)1423 $cm^{-1}$, (c) 1067 $cm^{-1}$, (d)1482 $cm^{-1}$ respectively.

**Supporting Note #6.** *Calculating area ratios of Adenine UV-SERS spectra*

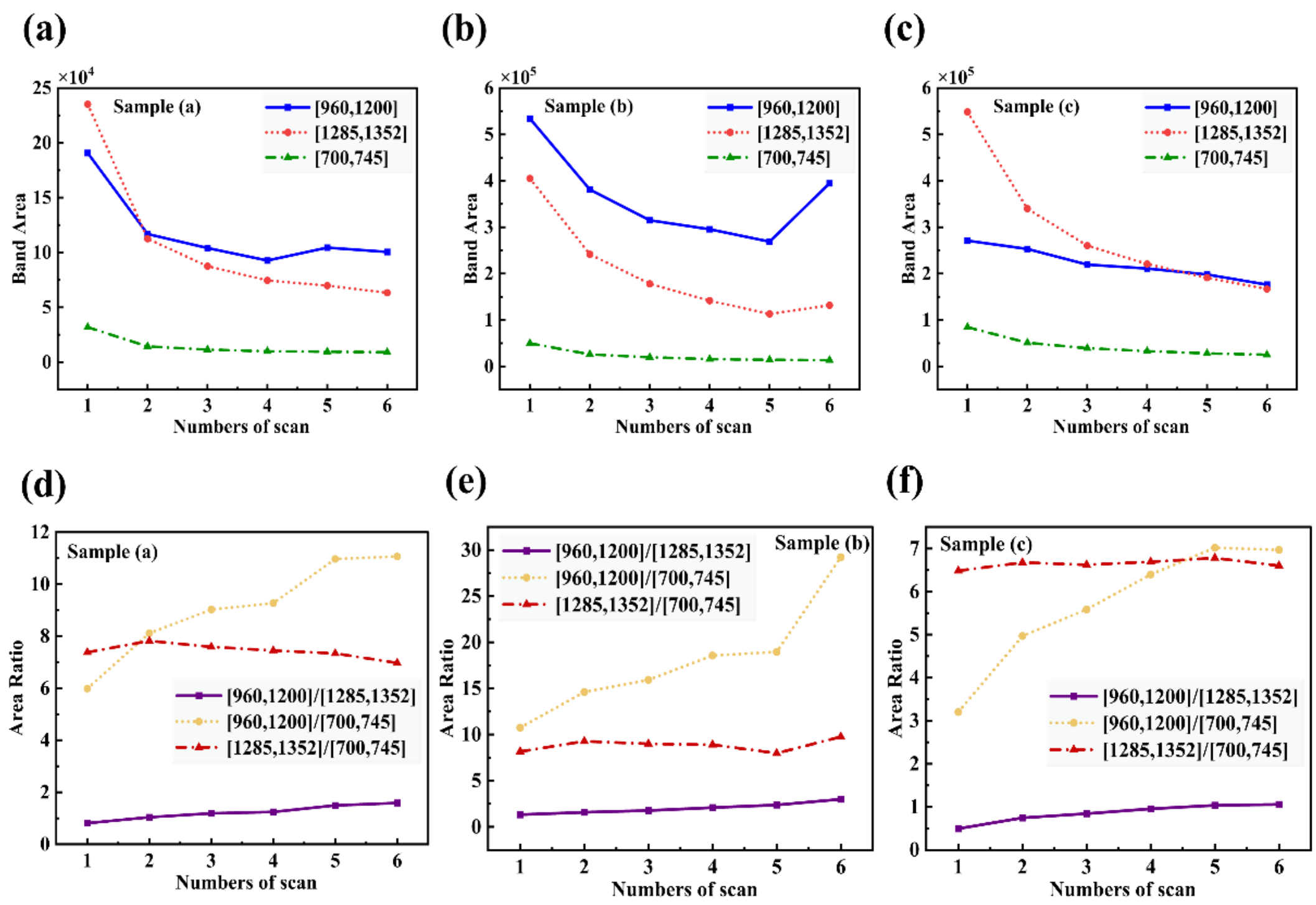

**Figure S8.** Calculating (a-c) the band area and (d-f) area ratio to different samples.

We calculated the band area corresponding to the wavenumber ranges of [700, 745], [960, 1200], and [1285, 1352] $cm^{-1}$ of adenine obtained the mean intensity across six continuous scans and the corresponding area ratio by using Python Jupyter, the results were reported in Fig. S8. For the Adenine UV-SERS peak area in the wavenumber range of [700, 745] $cm^{-1}$, the area decreases by approximately 50% from the first scan to the second scan. When the number of scans increases again, the peak area remains almost unchanged. This indicates that continuous scanning has little effect on the photodegradation of adenine around the wavenumber of 731 $cm^{-1}$. This initial reduction can be attributed to the rapid photodegradation of the most accessible and photoreactive adenine molecules under intense laser exposure during the first scan. The laser-induced breakdown of these molecules diminishes their Raman-active concentration, resulting in a substantial decline in the peak intensity. Once the susceptible adenine molecules are largely degraded, further scans encounter a stabilized population of less reactive or protected molecules, leading to minimal additional changes in the peak area. Additionally, the formation of a passivating layer of photodegradation products on the SERS-active substrate may shield the remaining adenine molecules from further laser-induced degradation. Consequently, the peak at 731 $cm^{-1}$ remains relatively constant, indicating that continuous scanning exerts limited further photodegradation effects on adenine within this specific wavenumber region. This stabilization highlights the initial vulnerability of adenine to photodegradation and the subsequent protective mechanisms that mitigate ongoing degradation under sustained scanning conditions. However, for the peak area corresponding to the wavenumber range of [1285, 1352] $cm^{-1}$, a gradual reduction is observed as the number of scans increases. With the addition of Rh NPs, this reduction process slows down, which is consistent with our previous results (DOI: 10.1039/d4ma00203b) and the data analyses in Fig. 3 in the main text. On the other hand, the peak areas in the wavenumber range of [960, 1200] $cm^{-1}$ generally decrease with the increasing number of scans, but no further patterns can be derived.

We performed a ratio analysis of the peak areas within the three wavenumber ranges, it can be seen from the line connected by red triangles in Fig. S8d-f (Ratio [1285,1352] / [700,745]) that the ratio of the "normal peak" corresponding to adenine remains basically constant. The line connected by purple circles (Ratio [960,1200] / [1285,1352]) shows a slow linear increase with the increasing number of scans. The line connected by yellow squares (Ratio [960,1200] / [700,745]) exhibits significant changes in the peak area ratios within the same range for different samples as the number of scans increases.

**Supporting Note #7.** *Peak deconvolution for UV-SERS spectra of Adenine*

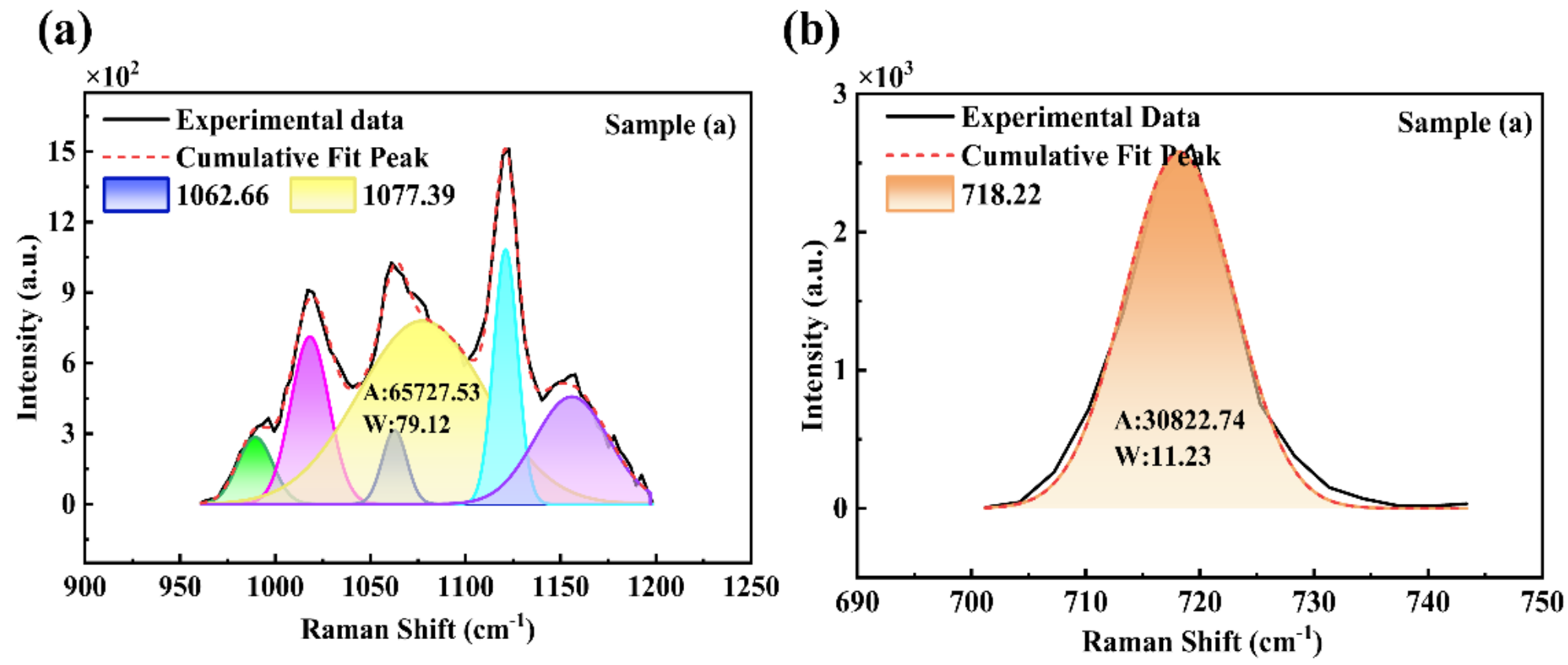

**Figure S9.** Fitting peak, corresponding peak area and FWHM of sample a ranging of (a) [960,1200]; and (b) [700,745]. A: Peak Area; W: Full width at half maximum (FWHM).

Fig. S9 reported the fitting results for sample a as an example to illustrate our reporting method by using Eq. (2) in the main text and listed the peak areas and full widths at half maximum (FWHM) for the fitted peaks of Sample a at 1077.39 cm$^{-1}$, and 718.22 cm$^{-1}$ (we labeled 731cm$^{-1}$ for the main characteristic adenine SERS peaks in the main text). Here, **A** represents the peak area, and **W** represents the full width with half maximum (FWHM). A similar treatment was also performed on the other two samples. As we can see in Fig. S9a, three fitted peaks appear in the 1000–1100 cm$^{-1}$ range, specifically at 1018 cm$^{-1}$, 1062.66 cm$^{-1}$, and 1077.39 cm$^{-1}$. These fitted peaks also correspond to the Raman characteristic peaks of adenine and its derivatives observed in the original spectra presented in our previous tests in Fig. S6. For the fitted peak at 1077.39 cm$^{-1}$, the full width at half maximum (FWHM) is 79.12 cm$^{-1}$, and the entire peak position encompasses the oxidation peak at 1048 cm$^{-1}$ reported in https://doi.org/10.1021/acs.jpclett.3c00619.

Since the area corresponding to the wavenumber range of [700-745] cm$^{-1}$ does not change significantly with an increasing number of scans, we selected the fitting peak at approximately 731 cm$^{-1}$ within this range as comparison. We then calculated the ratio of the peak area for each fitted peak to the peak area around 731 cm$^{-1}$ obtained from the same spectrum. The calculation results are shown in Fig. S10a-c, which report the variation of the peak area corresponding to the fitting peak across six consecutive scans. The green, red, and blue lines correspond to the peak area ratios at around 1075, 1020, and 1062 cm$^{-1}$ relative to 731 cm$^{-1}$, respectively. The results show the ratio of the fitted peak areas increases as the number of scans increases. When combined with the Rhodium NPs coverage percentage in **Table 1** in the main text, we can conclude that higher concentrations of Rh NPs promote greater oxidation effects, as evidenced by the steeper slopes observed at around 1075 cm$^{-1}$ fitted peak, which is indicative of enhanced oxidative interactions within the sample matrix. Specifically, sample b with 6.46% Rh NPs coverage exhibits the highest slopes across all ratios, suggesting enhanced oxidation facilitated by Rh NPs. However, sample c, despite having a higher Rh NPs coverage of 11.49%, shows moderated slope values, indicating potential stabilization effects at higher NP concentrations. Sample a, devoid of Rh NPs, demonstrates variable slope values, highlighting the critical role of Rh NPs in modulating oxidative processes.

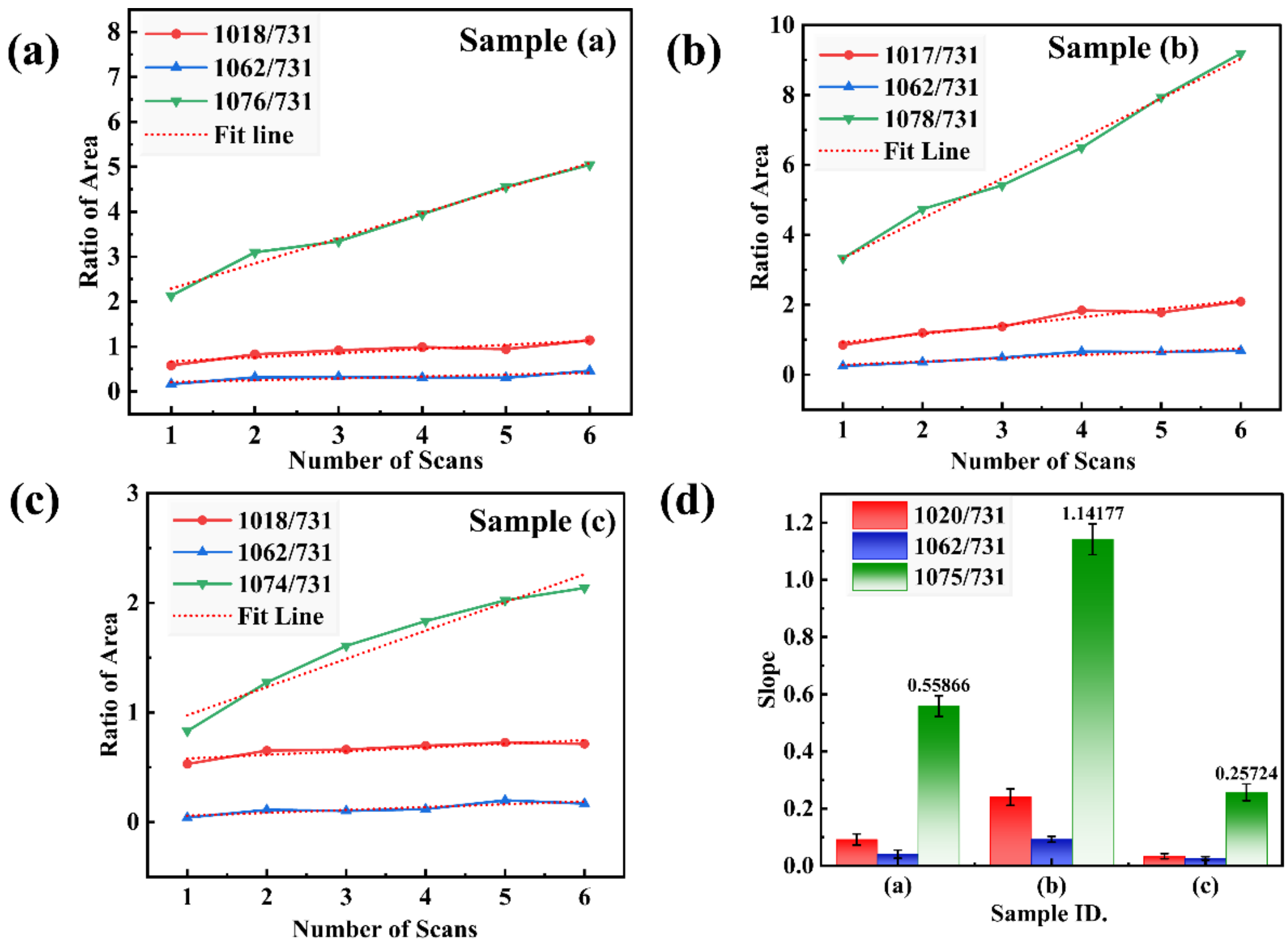

**Figure S10.** (a-c) Fitted peak ratios of adenine UV-SERS spectra and (d) slopes of linear fits for sample a-c. The fitting peak area ratios obtained from the SERS spectra for six consecutive scans across samples (a), (b) and (c), respectively.

Consequently, we derived the fitting peak area ratio for three samples, the red dashed lines indicate the linear fits applied to these ratios, which demonstrated substantial variation. Fig. S10(d) summarizes the slopes of these linear fits for each sample, along with their associated error bars, highlighting the reproducibility and consistency of the SERS measurements across different samples. It is noteworthy that, during the fitting process, a confidence interval of ±10 wavenumbers were set around the fitting peak at approximately 1075 $cm^{-1}$. The results show the peak area fitted at approximately 1075 $cm^{-1}$ varies significantly with an increasing number of scans. The data underscores the influence of Rh NPs coverage and scan repetitions on the oxidative dynamics of adenine, as reflected in the altered SERS peak area ratios within the 1000-1100 $cm^{-1}$ range.

**Supporting Note #9.** *UV-SERS spectra of bulk BSA under different laser excitation*

We test the bulk BSA on silica chip across different excitation wavelength, the result is reported in Fig. S11.

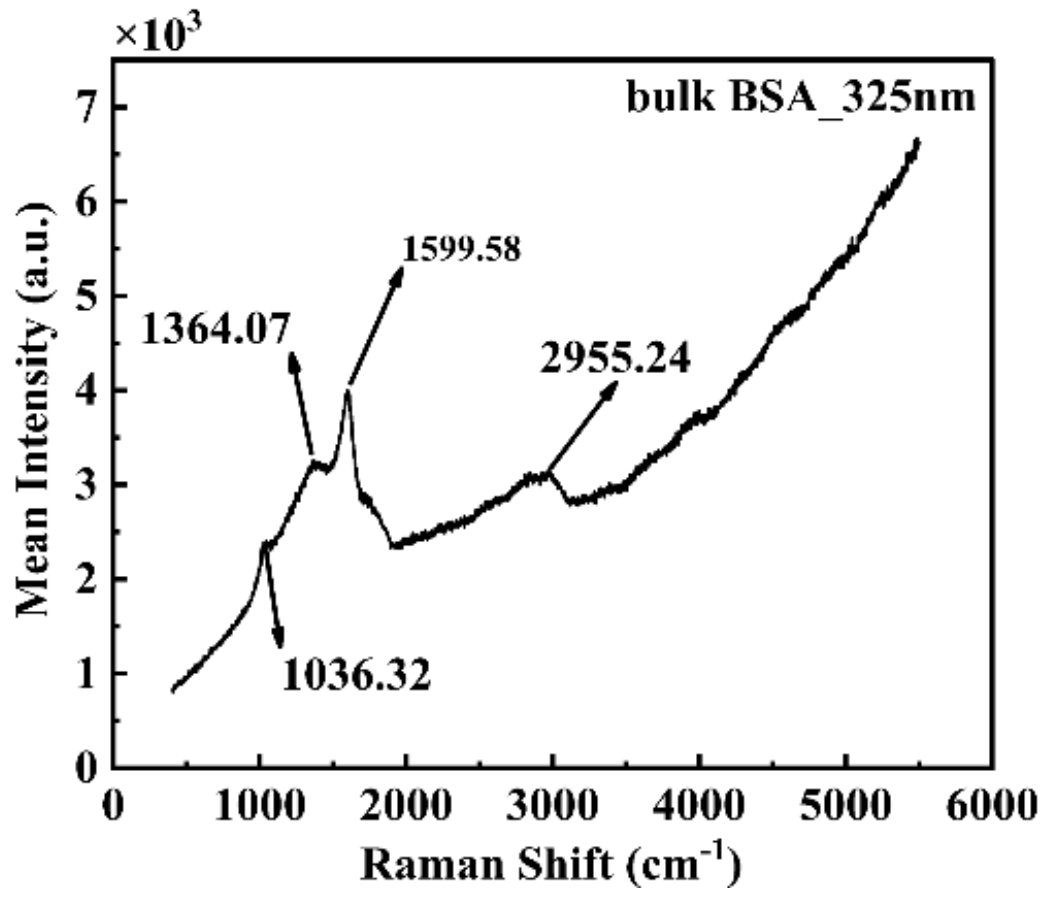

**Figure S11.** Raman spectra of bulk BSA - laser excitation wavelength at 325nm.